\documentclass[prl,twocolumn,amsmath,amssymb,floatfix,showpacs,superscriptaddress]{revtex4}

\usepackage{bm}

\usepackage{amssymb,amsmath}
\usepackage{graphics}
\usepackage{epstopdf}
\usepackage{epsfig}
\usepackage{color}
\usepackage{amsfonts}
\usepackage{bm}
\usepackage{amscd}
\usepackage{epsfig}
\usepackage{subfigure}
\usepackage{color}
\newcommand{\up}{\uparrow}
\newcommand{\down}{\downarrow}
\def\nab{{\mbox{\boldmath{$\nabla$}}}}

\newcommand{\figref}[1]{Fig.~\ref{#1}}

\begin{document}
\title{
Scale-dependent competing interactions:\\ sign reversal of the average persistent current  }
\author{H. Bary-Soroker}
\affiliation{Physics Department, Ben Gurion University,  Beer Sheva
84105, Israel}

\author{O. Entin-Wohlman}

\affiliation{Physics Department, Ben Gurion University,  Beer Sheva
84105, Israel}
\affiliation{Raymond and Beverly Sackler School of Physics and Astronomy, Tel Aviv University, Tel Aviv 69978, Israel}

\author{ Y. Imry}
\affiliation{Department of Condensed Matter Physics, Weizmann Institute of Science, Rehovot 76100, Israel}

\author{A. Aharony}
\affiliation{Physics Department, Ben Gurion University,  Beer Sheva
84105, Israel}
\affiliation{Raymond and Beverly Sackler School of Physics and Astronomy, Tel Aviv University, Tel Aviv 69978, Israel}

\date{\today}

\begin{abstract}

The interaction-induced orbital magnetic response of a nanoscale 
system, modeled by the persistent current in a ring geometry, is evaluated for a system which is a superconductor in the bulk.
The interplay of the renormalized Coulomb and Fr\"{o}hlich interactions is crucial. The diamagnetic response of the large superconductor may become paramagnetic when the finite-size-determined Thouless energy is larger than or on the order of the Debye energy. 

\end{abstract}

\pacs{73.23.Ra,74.25.N-,74.25.Bt}

\maketitle

\noindent{\bf Introduction.}
Renormalization, often accomplished using the renormalization-group  method, is one of the basic concepts in physics. It deals with the way various coupling constants (e.g.,  the electron charge or the Coulomb interaction) change as a function of the relevant  scale for  the given problem. 
The scale may be 
the resolution at which the system is examined, 
determined by its size and by the relevant energy for the process under consideration. Often, one knows the coupling constant's  ``bare value" 
at a less interesting (e.g., a very small or a very large) scale, and what is relevant for experiments is the value  at a different, ``physical" scale; the coupling  on the latter scale  is then used for the relevant physics \cite{fro}. 
Here we suggest a way to address experimentally the scale-dependence of the  renormalized interaction
by monitoring the magnetic response of a mesoscopic system.

\begin{figure}[thb]
\includegraphics[width=7.5cm]{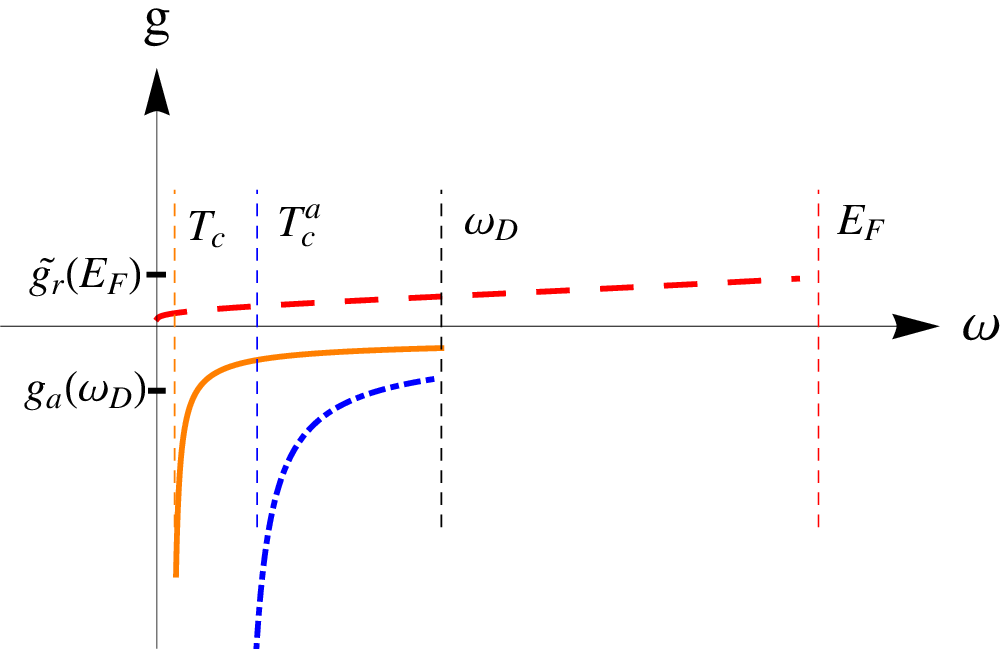}
\vskip 0.4cm
\includegraphics[width=9cm]{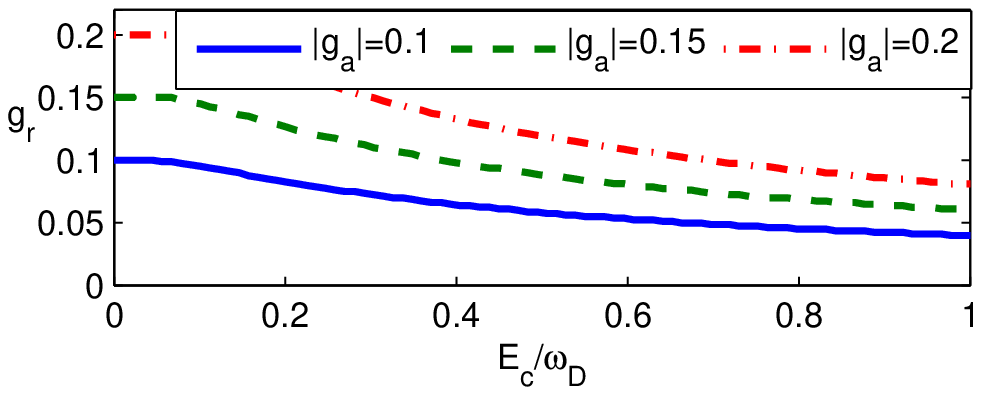}
\caption{Upper panel: the renormalized interactions, Eq. (\ref{RG}), as functions of the energy, for a metal which is a superconductor {\em in the bulk.} 
When only the attraction exists, its absolute value is renormalized up 
(with $\omega_{>}=\omega_{\rm D}$)
until it diverges at $T_{c}^{a}$ (the dash-dotted curve).
The repulsion (dashed curve) is renormalized down from its bare value at   $\omega_{>}=E_{\rm F}$. The total interaction starts at $\omega_{\rm D}$ from the bare value $g_{a}(\omega^{}_{\rm D})+g_{r}(\omega_{\rm D})$ and then its absolute value is renormalized up until it diverges at  the ``true" transition temperature $T_{c}$ (full curve). In a mesoscopic system the renormalization stops at the Thouless energy and therefore the relative location of $E_{c}$ {\it vs.} $\omega_{\rm D}$ and the ratio  $|g_{a}|/g_{r}({\rm max}\{E_{c}, \omega_{\rm D}\})$ will determine the sign of the effective interaction at $\omega_{\rm D}$. 
The figure is drawn for demonstration purposes with the unrealistic  parameter $E_{\rm F}/\omega_{\rm D}=2.5.$  
Lower panel:
the lines in the plane  $g_{r}-E_{c}/\omega_{\rm D}$ where the first harmonic of the current changes sign from 
paramagnetic (above the curve) to diamagnetic (below), for three values of the attraction $g_{a}$. Here and in all figures below, 
$T=\omega_{\rm D}/125
\simeq 2-4$K, and $E_{\rm F}/\omega^{}_{\rm D}=25$.}  \label{fig:phaseD}
\end{figure}

An important and particularly interesting application of this idea is in the theory of superconductivity.
There, two competing interactions exist: the repulsive Coulomb interaction,   starting on the large, microscopic energy scale--typically the Fermi energy/bandwidth $E_{\rm F}$, and the attractive phonon-induced interaction,  operative only below the much smaller Debye energy $\omega_{\rm D}$.
By integrating over thin shells in momentum (or energy) space \cite{MA}, one obtains  
the well-known variation of any coupling $g$, being repulsive or attractive,  from 
a high-energy scale $\omega_{>}$ 
to a low one $\omega_{<}$,   
\begin{align}
g(\omega^{}_{<})=\frac{g(\omega^{}_{>})}{1+g(\omega^{}_{>})\ln(\omega^{}_{>}/\omega^{}_{<})}\ .\label{RG}
\end{align} 
(We use 
$\hbar=c=k_{\rm B}=1$.)  
Notice that a repulsive/attractive interaction is ``renormalized downwards/upwards"  with decreasing energy scale.
What makes superconductivity possible is that at $\omega_{\rm D}$ the repulsion is much smaller than its value on the microscopic scale. 
At $\omega_{\rm D}$ the attraction may  win and then at lower energies the total interaction  increases in absolute value, until it diverges at some small scale, 
the conventional  ``$T_c$" of the given material.  The upper panel of Fig.   \ref{fig:phaseD} presents this qualitative  text-book picture. However, 
when the repulsive interaction wins at $\omega_{\rm D}$, the total interaction stays repulsive 
at all lower scales.
In this Letter we present and employ quantitative results for a   diffusive system  whose size $L$ exceeds the elastic mean-free path $\ell=v_{\rm F}\tau_{i}$ ($v_{\rm F}$ is the Fermi velocity and $\tau_{i}$ is the mean-free time).

An intriguing 
situation offered by mesoscopic superconductors, is the possibility to observe the ``running", scale-dependent  relevant interaction \cite{we}.  Obviously the renormalization has to be stopped at the scale corresponding to the system size,   which 
for many physical observables
is  
the Thouless energy 
which for diffusive systems is  $E_{c}=D/L^{2}$
($D=v_{\rm F}\ell /3$ is the diffusion coefficient) \cite{book}. Moreover, $E_{c}$ is also the energy scale for many mesoscopic phenomena.
Thus, by measuring   a quantity which depends on the interaction, as a function of $L$, {\em one may deduce the scale-dependence (``running" value) of that interaction.}
When that quantity  
is sensitive to the sign of the interaction (e.g.,  a linear function, up to corrections,  of the interaction, often resulting from perturbation theory),
one may  observe a rather dramatic change of the {\em sign} of the interaction-dependent term in that quantity!

A unique example for such a quantity 
is the orbital magnetic susceptibility of a mesoscopic/nanometric 
system, notably 
the persistent current \cite{BIL} flowing in a mesoscopic normal-metal ring in response to an Aharonov-Bohm  flux $\Phi$ through its opening. The magnetic moment induced by very small fluxes due to these currents is expected to be of the same order of magnitude as that of a small disc made of the same material, and having  similar radius and thickness, in response to the same flux.

Moreover, it is well known \cite{book,AE}  that the impurity ensemble-averaged persistent current for noninteracting electrons is smaller by typically two orders of magnitude than the interaction-dominated terms. Therefore 
the latter represents rather well the whole magnetic response of the electrons and its measurement as a function of $L$ will amount to monitoring the renormalized interaction.  This picture
is borne out by our  calculations. 
We implement      
the ideas of Refs. \onlinecite{MA} 
in the derivation of the
superconductive fluctuation-induced partition function, and   obtain the effect of the two competing interactions in the Cooper channel. We use this result to
study the  average scale-dependent persistent current  of a large ensemble of metallic  rings in the presence of these  interactions. 

The
renormalization of the interactions has  to be carried out down to an energy scale roughly given by max$\{T,E_{c}\}$, where $T$ is the temperature.  
We denote by $g_{a}$  the attraction and by  $g_{r}=[\widetilde{g}^{-1}_{r}+\ln (E^{}_{\rm F}/\omega^{}_{\rm D})]^{-1}$ the repulsion, both at $\omega_{\rm D}$,
such that for  $|g_{a}|>g_{r}$, the material in its bulk form is a superconductor.
When $E_{c}>T$ then
depending on the ratios $E_{c}/\omega_{\rm D}$   and $|g_{a}|/g_{r}({\rm max}\{E_{c}, \omega_{\rm D}\})$
the resultant interaction may be repulsive or attractive, and  the response may be paramagnetic or diamagnetic. These two types of behavior are reflected by the sign of the average current at small fluxes. 
The lower panel of  \figref{fig:phaseD}  presents a ``phase diagram"  showing   the regions where  the first harmonic of the current   is paramagnetic or diamagnetic, as a function of $E_{c}/\omega_{\rm D}$. 
Note in particular the existence of a region where although the attraction  $|g_{a}|$ exceeds the repulsion $g_{r}$,  the current is {\em paramagnetic}, contrary to the usual expectation \cite{AE}.
The Thouless energy can be increased by making the ring smaller, and/or 
less disordered.

\noindent{\bf Renormalizing  the two  interactions.}
The Hamiltonian we consider consists of a single-particle part, $H_{0}({\bf r})=(-i\nab-e{\bf A})^{2}/(2m)-\mu+u({\bf r})$, and a term describing {\em local} repulsion and attraction, of coupling constants $\widetilde{g}_{r}\equiv\lambda_{r} {\cal N}/V$ and $g_{a}\equiv\lambda_{a}{\cal N}/V$, respectively,
\begin{align}
{\cal H}&=\int d{\bf r} \Bigl (\sum_{\sigma}\psi^{\dagger}_{\sigma}({\bf r})H^{}_{0}({\bf r})\psi^{}_{\sigma}({\bf r})\nonumber\\
&+(\lambda^{}_{r}-|\lambda^{}_{a}|)
\psi^{\dagger}_{\up}({\bf r})\psi^{\dagger}_{\down}({\bf r})\psi^{}_{\down}({\bf r})\psi^{}_{\up}({\bf r})\Bigr )
\ ,\label{HAM}
\end{align}
where 
${\cal N}$ is the density of states,   $V$  the system volume, 
and
$\psi^{}_{\sigma}$ ($\psi^{\dagger}_{\sigma}$) destroys (creates) an electron of spin  component  $\sigma$ at ${\bf r}$.  
Here, $\widetilde{g}_{r}$ is the bare repulsion, $\mu$ is the chemical potential,  
$u({\bf r})$ is the disorder potential due to nonmagnetic impurities, and ${\bf A}$ is the  vector potential,
with $A=\Phi /L$,  $\Phi$ giving rise to a persistent current.
As the magnetic fields needed to produce  such currents are rather small, the Zeeman interaction will be ignored.

The partition function, ${\cal Z}$, corresponding to the Hamiltonian (\ref{HAM}) is calculated by the method of Feynman path integrals, combined with the Grassmann algebra of many-body fermionic coherent states \cite{AS}. Introducing the bosonic fields $\phi({\bf r},\tau)$ and $\Delta ({\bf r},\tau)$
via
{\em two} Hubbard-Stratonovich transformations, {\em both} carried out in the Cooper channel  \cite{SM}, 
one is able to integrate over the fermionic variables; the result is then expanded 
to second order in the bosonic fields \cite{COM3}
\begin{align}
\label{Zm}
&{\cal Z}/{\cal Z}_{0}= \int D(\Delta,\Delta^{\ast}_{}) \int D(\phi,\phi^{\ast}_{})
\exp[-\widetilde {\cal S}]\ ,
\end{align}
with the action $\widetilde{\cal S}$ given by
\begin{align}
&\widetilde {\cal S}=\sum_{{\bf q},\nu} \Bigl (\beta {\cal N} |\Delta^{}_{{\bf q},\nu}|^{2}/|g^{}_{a}|
+\beta {\cal N}|\phi_{{\bf q},\nu}|^{2}/\widetilde{g}^{}_{r} \nonumber\\& -\left(
\Delta_{{\bf q},\nu}+i\phi^{}_{{\bf q},\nu}\right)\left(
\Delta_{{\bf q},\nu}^{\ast}+i\phi_{{\bf q},\nu}^{\ast}\right)
 \widetilde\Pi_{{\bf q},\nu} \Bigr )\ ,\label{Am}
\end{align}
and ${\cal Z}_{0}$ denoting the partition function of  noninteracting electrons
($\beta =1/T$). 
Equation (\ref{Zm}) represents the partition function due to the pair-field fluctuations, with the
action  $\widetilde{\cal S}$,   a sum over the wavevectors ${\bf q}$ and the bosonic Matsubara frequencies $\nu=2\pi n T$,  written in terms of the polarization $\widetilde{\Pi}$ 
averaged over the impurity disorder  \cite{AGD}. It is this object in which the different renormalization of the competing interactions manifests itself. It is important to note that both bosonic fields, $\Delta $ and $\phi$, pertain to {\em pair-field fluctuations}; the first resulting from the decoupling of  the bare attractive interaction, while the second arises from the decoupling of bare repulsive one [i.e., the Thomas-Fermi screened Coulomb potential, denoted $\lambda_{r}$ in Eq. (\ref{HAM})].

In a diffusive system, the disorder-averaged polarization
is given as a sum over fermionic Matsubara frequencies, $\omega =\pi T(2n+1)$,
\begin{align}
\widetilde{\Pi}^{}_{{\bf q},\nu}
&= {\cal N}\sum_{\omega}
\frac{2\pi\Theta[\omega (\omega +\nu)]}{|2\omega+\nu|+Dq^{2}}=\frac{ {\cal N}}{T}\sum_{n}^{} \frac{1}{ n+ F^{}_{{\bf q},\nu} }  \ ,\nonumber\\
\label{F}
&\hspace{1cm}F^{}_{{\bf q},\nu}=0.5+(Dq^2+|\nu|)/(4\pi T)\  ,
\end{align}
where $\Theta $ is the Heaviside function.
The first of Eqs. (\ref{F}) is valid for  $|2\omega+\nu|\ll\tau_{i}^{-1}$ and  $q\ell\ll 1$.
It represents a divergent sum
and therefore a cutoff is needed.  
The cutoff energy is determined by the relevant interaction. 
The sums in the terms of Eq. (\ref{Am})  involving $\Delta$ are bounded by $\omega_{\rm D}$ and therefore are cut off at 
$n_{\max }=\omega^{}_{\rm D}/(2\pi T)-1$; these sums are denoted 
$\Pi^{\omega_{\rm D}}$.   The sum multiplying $|\phi|^2$ is cut off by $E^{}_{{\rm F}}$, i.e., $n_{\max }=E^{}_{\rm F}/(2\pi T)-1$; this sum is denoted by $\Pi^{E_{\rm F}}$. (Note that in the diffusive regime both cutoffs are smaller than $\tau^{-1}_{i}$, i.e., $\omega^{}_{\rm D}\tau^{}_{i}<E^{}_{\rm F}\tau^{}_{i}<1$.) Therefore, there appear two polarizations, 
the one cut off  by $\omega_{\rm D}$ and the other by $E_{\rm F}$ 
\begin{align}
\frac{T}{{\cal N}}\Pi^{\omega^{}_{\rm D}( E^{}_{\rm F})}_{{\bf q},\nu}=
\Psi\Bigl (\frac{\omega^{}_{\rm D}(E^{}_{\rm F})}{2\pi T}+F^{}_{{\bf q},\nu}\Bigr )-
\Psi\Bigl (F^{}_{{\bf q},\nu}\Bigr )\ ,\label{PID}
\end{align}
where $\Psi$ is the digamma function. 

Performing the integration over the fields $\phi$ yields 
\begin{align}
\label{z5a}
\frac{\cal Z}{{\cal Z}_0}= &\prod_{{\bf q},\nu}  \int D(\Delta^{}_{{\bf q},\nu},\Delta^{\ast}_{{\bf q},\nu})
\exp\Bigl [- \Bigl (\frac{ 1}{g^{}_{r}({\bf q},\nu)
}
-\frac{ 1}{|g^{}_{a}|}\Bigr )\nonumber\\
&\times\Bigl (\frac{\beta {\cal N}}{|g^{}_{a}|-g^{}_{r}({\bf q},\nu)}-
\Pi^{\omega_{\rm D}}_{{\bf q},\nu}\Bigr ) |\Delta^{}_{{\bf q},\nu}|^2 \Bigr ] \ , 
\end{align}
where $g_r({\bf q},\nu)$ is the wave vector- and frequency-dependent repulsion, renormalized from $E_{\rm F}$ down to $\omega_{\rm D}$   
\begin{align}
\label{gtil}
g^{-1}_{r}({\bf q},\nu)\
&=\widetilde{g}^{-1}_{r}+\Psi\Bigl (\frac{E^{}_{\rm F}}{2\pi T}+F^{}_{{\bf q},\nu}\Bigr )-\Psi\Bigl (\frac{\omega^{}_{\rm D}}{2\pi T}+F^{}_{{\bf q},\nu}\Bigr )\ .
\end{align}
In the limit $\omega^{}_{\rm D}\gg\{Dq^{2},T\}$
one may neglect $F_{{\bf q},\nu}$ in the arguments of the digamma functions    
and retrieve  (exploiting the asymptotic form of $\Psi$) the well-known result
$g^{-1}_{r}=\widetilde{g}^{-1}_{r}+\ln (E^{}_{\rm F}/\omega^{}_{\rm D})$.
In a finite ring, however, the smallest wavevector is of the order $1/L$ and 
therefore at low temperatures this procedure  is valid
only when the Thouless energy is smaller than the Debye energy, i.e., for large enough (and/or clean enough) rings. 
Carrying out the integrations  over the fields $\Delta$ in Eq. (\ref{z5a})  yields
\begin{align}
{\cal Z}/{\cal Z}^{}_{0}=\prod_{{\bf q},\nu}a^{-1}_{{\bf q},\nu}\ ,\label{ZP}
\end{align}
with  $a_{{\bf q},\nu}=a(F=F_{{\bf q},\nu})$
and
\begin{align}\label{a}
& a(F)=\frac{
1}{|g^{}_{a}|-g^{}_{r}}-
\Psi\left(F+\frac{\omega^{}_{\rm D}}{2\pi T}\right) +\Psi\left(F\right)
\nonumber\\&+\frac{ g^{}_{r}|g^{}_{a} |}{ |g_{a}|-g_{r}}
\left[\frac{
1}{|g^{}_{a}|}-
\Psi\left(F+\frac{\omega^{}_{\rm D}}{2\pi T}\right)
+\Psi\left(F\right)\right]\\&\times\left[ \Psi\left(F+\frac{E^{}_{\rm F}}{2\pi T}\right)-\ln\left(\frac{ E^{}_{\rm F}}{\omega^{}_{\rm D}}\right)
-\Psi\left(F+\frac{\omega^{}_{\rm D}}{2\pi T}\right)\right]\ .\nonumber
\end{align}
Equation (\ref{ZP}) is our result for the superconductive fluctuation-induced partition function. It is based on the 
renormalized repulsion, Eq. (\ref{gtil}).

\noindent{\bf 
The fluctuation-induced average persistent current.}
The thermodynamic persistent current is
obtained by differentiating the free energy with respect to the dimensionless flux,  $\varphi=\Phi/\Phi_{0}$,  i.e., $  I=(eT)/(2\pi)\partial(\ln {\cal Z})/\partial\varphi  $
where $\Phi_{0}=h/e$ is the flux quantum. 
In the ring geometry, the flux shifts the longitudinal
component, $q_{\parallel}$, of the wavevector ${\bf q}$, $q^{}_{\parallel}\equiv q^{}_{n}=2\pi(n+2\varphi )/L$
where $n$ is an integer \cite{we}. 
Hence (exploiting the Poisson summation formula)
\begin{align}\label{harmonics}
I= \sum_{m=1}^{\infty}\sin(4\pi m\varphi) I^{}_m\ ,
\end{align}
where \cite{we}
\begin{align}\label{current}
I^{}_m=[(2ieT)/\pi]\sum_\nu\int_{-\infty}^\infty dx  e^{2\pi i x} \partial\ln[a^{}_{x,m,\nu}]/\partial x\ ,
\end{align}
with $a_{x,m,\nu}$   [Eq. (\ref{a})]   given in terms of   $      F_{x,m,\nu}=0.5+|\nu|/(4\pi T)+ (\pi E^{}_c x^2)/(m^2 T)$            
[see Eq. (\ref{F})].
The integral in Eq. (\ref{current}) is handled by a contour integration, and the sign of the result {\em which determines  the magnetic response},   depends on a subtle balance between the contribution of the poles and that of the zeros \cite{we}.

It is customary in studies of persistent currents to concentrate on the lowest harmonics, notably on $I_{m=1}$, since 
higher harmonics are exponentially smaller compared to the lowest ones  and are also more sensitive to inelastic scattering (dephasing) and pair-breakers (e.g., magnetic impurities)   \cite{we}.
However, when 
$T<E_{c}$
the relative magnitude of  harmonics higher than the first one increases, and their contribution should not be discarded.

When only one of the two competing interactions is accounted for, the nature of the magnetic response is dictated by the sign of that interaction,  being diamagnetic for attraction and paramagnetic for repulsion \cite{AE}.  When both interactions are present the picture is changed, as is exhibited in the lower panel of Fig. \ref{fig:phaseD}, obtained by numerically calculating  $I_{m=1}$   
as a function of the Thouless energy $E_{c}$. 
The sign of $I_{m=1}$
depends on the ratio $E_c/\omega_{\rm D}$. As the latter increases,   this sign 
may change from  negative to  positive. On the other hand, when $g^{}_{r}>|g^{}_{a}|$ (and the bulk material will not become a superconductor at a finite temperature) the current remains paramagnetic for all values of $E_{c}/\omega_{\rm D}$.

\begin{figure}[hbtp]
\vspace{-0.5cm}
\includegraphics[width=9cm]{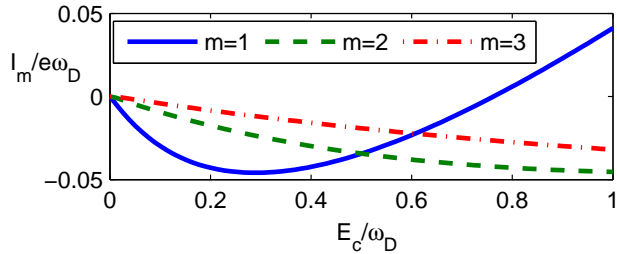}
\caption{$I_m(E_c)$ for $m=1$, 2, and 3 (see the legend), with    $|g^{}_{a}|=0.2$ and  $g^{}_{r}=0.1$.}
\label{fig:I_m}
\end{figure}

\begin{figure}[hbtp]
\includegraphics[width=8.5cm]{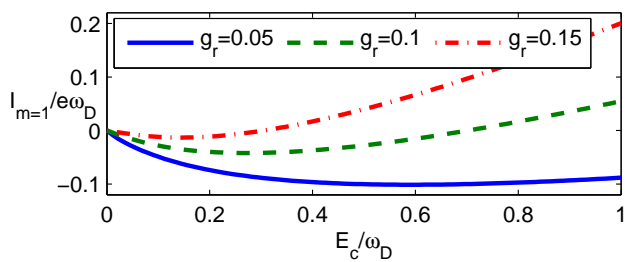}
\vspace{-0.35cm}
\caption{$I_{m=1}$ 
for three values of the repulsion and $|g_{a}|=0.2$. }
\label{fig:I_Ec_several_mu}
\end{figure}

\begin{figure}[hbtp]
\includegraphics[width=7.5cm]{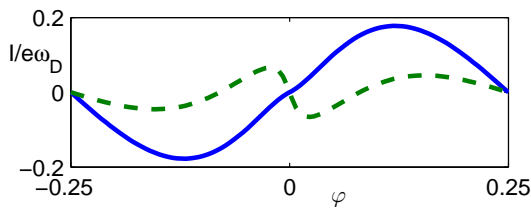}
\vspace{-0.35cm}
\caption{ The current as a function of $\varphi$. Solid (dashed) curve is for $g^{}_{r}=0.17$ ($g_{r}=0.1$), with $|g_a|=0.2$ and $E_c/\omega_{\rm D}=0.76$. }  
\label{fig:I_phi}
\end{figure}

As mentioned, it is  when $T>E_{c}$ that  the first harmonic represents faithfully the persistent current. 
Figure \ref
{fig:I_m} depicts the first three harmonics of the current, as functions of  the Thouless energy (both normalized by $\omega_{\rm D}$).
The figure indicates that
the value of $E_c$ at which the  sign of $I_m$ is reversed increases with $m$. 
Figure 
\ref{fig:I_Ec_several_mu} demonstrates the dependence of $I_{m=1}$ on the strength of the repulsion.    As it increases the $E_{c}/\omega_{\rm D}-$range in which the response is diamagnetic diminishes, and so does also the maximal value of the diamagnetic current. 
In the paramagnetic-response regime the amplitude of the current increases with $E_c$ \cite{book}. Figure \ref{fig:I_phi}  shows the flux dependence of the current, for two distinct values of $g_{r}$, and displays the possibility for a  sign reversal of the magnetic response at low flux. 
One notes in Fig.  \ref{fig:I_m} that at $E_{c}/\omega_{\rm D}=0.76$ $I_{m=1}$ is small compared to $I_{m=2,3}$; the dashed curve in  Fig. \ref{fig:I_phi} 
reflects the dominance of the higher harmonics over the first one for relatively small $g_{r}$. For a higher value  (solid line there) 
the first harmonic is dominant.


\noindent{\bf
Discussion.}
We have obtained the   wavevector- and frequency-dependent renormalized  repulsion, Eq. (\ref{gtil}). 
The two cutoffs on the bare attraction and repulsion, $\omega^{}_{\rm D}$ and  $E_{\rm F}$, respectively, conspire to determine the sign of the magnetic response. Whereas an overall total attraction suffices to lead to a superconducting phase in the bulk material, we find that it does  not ensure a diamagnetic response of the mesoscopic system. The reason being that in mesoscopic rings  the orbital  magnetic response owes its very existence to a finite Thouless energy. Therefore the latter, when large enough can cause the response to be paramagnetic, albeit the strength of the attractive interaction. As the Thouless energy may be controlled experimentally, one may hope that the prediction made in this paper will be put to an experimental test.  Relatively large values of persistent currents may be achieved in molecular systems, and small discs are expected to behave similarly at small fluxes.
Finally we remark that there may already be experimental indications
to the validity of our prediction. Reich {\it et al.} \cite{REICH} found that thin enough gold films are paramagnetic. This may well be due to a small grain structure.  


\begin{acknowledgments}
We are indebted to  D. Gross, Y. Oreg, and J. Feinberg for instructive discussions. AA and OEW acknowledge the support of the Albert Einstein Minerva Center for Theoretical Physics, Weizmann Institute of Science. This work was supported by the BMBF within the DIP program, BSF, ISF and its Converging Technologies Program.
\end{acknowledgments}

\end{document}